\documentclass[aps,prd,twocolumn,superscriptaddress,groupedaddress,nofootinbib]{revtex4-1}  

\usepackage{graphicx} 
 \usepackage{amsmath,amssymb,amsthm,dsfont}
\usepackage[colorlinks=true,citecolor=blue,linkcolor=blue,urlcolor=blue]{hyperref}

\usepackage{xcolor}
\usepackage{epsfig} 
\input{epsf}
\usepackage{natbib} 

\usepackage{longtable}
\usepackage{graphics}  
\usepackage{dcolumn}   
\usepackage{bm}        
\usepackage{amssymb}   

\def\nn{\nonumber}
\def\be{\begin{equation}}
\def\ee{\end{equation}}
\def\bea{\begin{eqnarray}}
\def\eea{\end{eqnarray}}

\def\ms{\overline{\rm MS}}

\def\ga{\gamma}
\def\ga0{\gamma_0}
\def\ds{\displaystyle}

\begin{document}

\title{All order resummed leading and next-to-leading soft modes of dense QCD pressure}

\author{Lo\"{\i}c Fernandez} 

\author{Jean-Lo\"{\i}c Kneur}
\affiliation{Laboratoire Charles Coulomb (L2C), UMR 5221 CNRS-Universit\'e Montpellier, 34095 Montpellier, France}

\begin{abstract}
The cold and dense QCD equation of state (EoS) at high baryon chemical potential $\mu_B$ involves
at order $\alpha_s^2$ 
an all-loop summation of the soft mode $m_E\sim \alpha_s^{1/2} \mu_B$ contributions. 
Recently, the complete soft contributions at order $\alpha_s^3$ were calculated, 
using the hard thermal loop (HTL) formalism.
By identifying {\em massive} renormalization group (RG) properties within HTL, 
we resum to all orders $\alpha_s^p, p\ge 3$ the leading and next-to-leading logarithmic soft contributions.  
We obtain compact analytical expressions, that show {\em visible}
deviations from the state-of-the art results, and noticeably reduced residual scale dependence.
Our results should help to reduce uncertainties in extending the EoS in the intermediate $\mu_B$ regime, 
relevant in particular for the phenomenology of neutron stars.  
\end{abstract}
%
\maketitle
{\em Introduction:}
At sufficiently high temperature or density, the asymptotic freedom 
property of quantum chromodynamics (QCD) naively provides a weak coupling perturbation theory (PT) to address
the quark-gluon plasma physics. However, severe infrared (IR) divergences spoil a
naive PT approach, giving poorly convergent results at successive orders, unless at extremely high
temperatures and/or densities
(see {\it e.g.} \cite{Trev} for reviews). 
In contrast, the nowadays powerful lattice simulations 
(LS) offer an alternative numerical nonperturbative ab initio solution. 
So far LS have been very successful in the description of the QCD crossover transition at 
finite temperatures and near vanishing baryonic 
densities, with results~\cite{lattice} also relevant to confront the experimental 
data from  heavy ion collisions in this specific region of the phase diagram.   
However, the notorious sign problem\cite{sign} prevents simulations at high densities, equivalently high chemical potential 
$\mu_B$, to explore the more complete QCD phase diagram and in particular at $\mu_B$ values pertinent to  
the physics of neutron stars\cite{NSrev}. Alternatively, 
more analytical approaches to resum thermal and in-medium PT have been
developed and refined over the years (see e.g. \cite{Trev,HTLpt,HTLpt2g,HTLptDense3L,HTLrev2020,EoSFuFu}),
improving the bad convergence generically observed even at moderate coupling values. More recently, a 
RG resummation approach giving sizeably improved renormalization 
scale uncertainties was developed\cite{rgopt_cold,rgopt_hot}.\\
In the following we focus on cold and dense QCD, $T=0,\mu_B \ne 0$, which implies a number of simplifications.
For strongly coupled matter at high baryonic density, the dynamical screening of color charges manifests in 
a screening mass defining a soft scale $m_E\sim \sqrt{\alpha_s} \mu_B \ll \mu_B$, 
where $\alpha_s$ is the QCD coupling.
This is linked to IR divergences in the naive perturbative pressure,
to be appropriately resummed, resulting in nonanalytic $\ln \alpha_s$ dependences  
in the perturbative expansion. This phenomenon occurs first at order $\alpha_s^2$, giving 
a contribution $\alpha_s^2 \ln \alpha_s$ 
established for massless quarks long ago\cite{soft1}. The pressure was extended much later to massive 
quarks\cite{Fraga:2004gz,LaineSchroder,coldQM2010}, but for a very long time no higher order was available. 
These soft terms are not the full contributions at $\alpha_s^{p\ge 2}$ orders, and should be completed
by hard contributions $\mu^4 \alpha_s^p$ calculable from standard PT, known exactly 
to date up to order $\alpha_s^2$\cite{soft1,pQCDmu4L}.
Yet, the soft terms constitute a well-defined subset, relevant for the convergence of the 
weak coupling expansion. The combined hard and soft contributions exhibit sizeable residual dependence 
in the (arbitrary) renormalization scale (although less severe than for thermal QCD), leading 
to systematic uncertainties.
It is therefore crucial to push further the weak coupling expansion, 
as one expects that higher perturbative orders may reduce uncertainties in the EoS, in particular in a regime 
presumably relevant to the physics of neutron stars\cite{NSrev}.
Given the abundant new data on compact stellar objects from astrophysics,
and the rapidly developing interplay between gravitational wave physics,  
QCD and nuclear calculations, 
it is timely to try to further tighten the existing gap
in the moderate $\mu_B$ regime,
between the reliable perturbative QCD at high $\mu_B$ and reliable EoS in the low $\mu_B$ nuclear regime.
Recently, the complete soft terms at the next order,  
$\alpha_s^3 \ln^p(m_E), 0\le p\le 2$, were obtained in \cite{Gorda:2018gpy,Gorda:2021kme,Gorda:2021znl},
from involved calculations using the hard thermal loop (HTL) formalism to unprecedented $m_E^4 \alpha_s$ order.\\ 
Our main purpose in this letter is to calculate and resum the soft leading logarithms (LL) and next-to-leading logarithms (NLL) 
to {\em all} $\alpha_s^p$ orders, $p\ge 3$. As particular case it provides an independent simpler derivation
of the first LL term $\alpha_s^3\ln^2\alpha_s$, consistent with \cite{Gorda:2021kme}. 

\begin{figure}[h!]
\centerline{ \epsfig{file=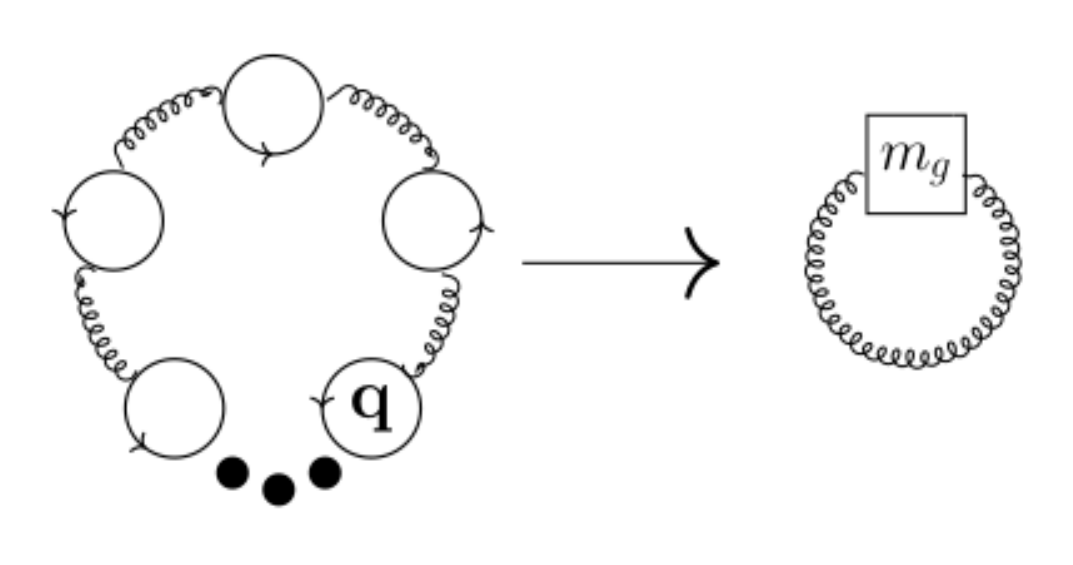,width=0.5\linewidth,angle=0}}
\caption{Connection between Ring sum and one-loop HTL.}
\label{fig:ringtoHTL}
\end{figure}

{\em Cold quark matter state-of-the-art pressure:}
For $N_f$ massless quarks with $\mu_q=\mu\equiv \mu_B/3$ 
the presently known quark matter weak expansion pressure reads
\bea
\ds
&P^{cqm} =  P_f\left[ 1 - \frac{2}{\pi} \alpha_s -\frac{N_f}{\pi^2} \alpha_s^2 \ln \alpha_s  -0.874355\, \alpha_s^2 \right. \nn \\
& \left. -2d_A\frac{(11N_c-2N_f)}{3(4\pi)^2} \ln (\frac{M_h}{\mu})\: \alpha_s^2 \right] + P^{soft}_{\alpha_s^3} ,
\label{Pas2}
\eea
with $d_A=N_c^2-1$ ($N_c=3$), and other terms specified and commented below.
The leading order (LO) massless quark loop gives the free gas pressure
$P_f= N_c N_f \mu^4/(12\pi^2)$,
and next-to-leading order (NLO) gluon exchange gives the ${\cal O}(\alpha_s)$ term. 
At $\alpha_s^2$ order  
the three-loop graphs in standard perturbation give\cite{soft1,pQCDmu4L} 
the hard contributions $\sim \alpha_s^2\mu^4/P_f$ 
in Eq.(\ref{Pas2}), also involving an arbitrary renormalization scale $M_h$.
The nonanalytic $\alpha_s^2 \ln \alpha_s$ term
arises from resumming\cite{soft1} the set of all order soft contributions, the 
``ring'' graphs (left in Fig.\ref{fig:ringtoHTL}). 
A well-known important feature is that the all-loop summation gets rid of initially IR divergences, 
a mechanism related to the dynamical gluon screening mass.
At $T=0, \mu \ne 0$, $m_E$ is obtained from the sole quark-loop contribution to the self-energy:
\be 
m^2_E = 2 \frac{\alpha_s}{\pi} \sum_f \mu_f^2 = 2 \frac{\alpha_s}{\pi} N_f \mu^2 .
\label{mEdef}
\ee
Accordingly, in a rough picture, the all-loop soft mode summation can be essentially obtained from a
one-loop calculation, within an alternative framework
with a massive gluon (the graph on the right in Fig.~\ref{fig:ringtoHTL}),
giving a contribution $P\sim m_E^4 \ln m_E \sim \mu^4 \alpha_s^2 \ln \alpha_s$.
This picture is rigorously embedded in the HTL formalism\cite{HTLbasic,HTLptDense3L}, 
that essentially provides a gauge invariant effective field theory (EFT) 
consistently including all HTL
contributions with dressed momentum-dependent self-energies and vertices,
and involving the screening gluon mass.
The $\alpha_s^3$ last contribution in Eq.(\ref{Pas2}) was 
calculated\cite{Gorda:2018gpy,Gorda:2021kme} from 
HTL graphs with only gluons, NLO corrections to the right graph in Fig.\ref{fig:ringtoHTL}. 
The $m_E^4 \alpha_s \ln^2(m_E)\sim \alpha_s^3 \ln^2 \alpha_s$ was first obtained
in \cite{Gorda:2018gpy},
and recently the remnant soft terms were completed,  
obtaining\cite{Gorda:2021kme,Gorda:2021znl}:
\bea
&& P^{\rm soft}_{\alpha_s^3}  =\frac{N_c d_A \alpha_s m^4_E}{(8\pi)^2} 
\left(\frac{p_{-2}}{4\epsilon^2}  +\frac{p_{-1}-2p_{-2} \ln \frac{m_E}{M_s}}{2\epsilon} \right. \nn \\
&&\left. +2p_{-2} \ln^2 \frac{m_E}{M_s} -2 p_{-1} \ln \frac{m_E}{M_s} +p_0 \right), 
\label{Psoftg3}
\eea
\be
p_{-2} = \frac{11}{6\pi},\;\;
p_{-1} \simeq 1.50731(19), \;\;
p_0 \simeq 2.2125(9) .
\label{p210}
\ee
All the 
$\mu$-dependence arises solely from the screening electrostatic mass, 
$m_E$ from Eq.(\ref{mEdef}).
Notice in Eqs.(\ref{Pas2})(\ref{Psoftg3}), besides $M_h$, the different 
scale $M_s$ introduced in \cite{Gorda:2021znl}\footnote{Our
$M_s$ scale corresponds to $\Lambda_h$ in \cite{Gorda:2021znl}.},
for the soft sector. As explained in \cite{Gorda:2021kme,Gorda:2021znl} the presently unknown $\alpha_s^3$
hard (and mixed soft-hard) contributions are expected to cancel the remnant UV divergences and 
soft $\ln^p (m_E/M_s)$ 
in (\ref{Psoftg3}), to let only $\ln^p \alpha_s$ and $\ln^p (M_h/\mu)$ terms ($p=1,2$). 
In absence of such explicit cancellations, since the soft terms can be treated as 
a separate $m_E$-dependent sector, to avoid large logarithms it appears sensible to choose\cite{Gorda:2021znl} $M_h\sim {\cal O}(\mu)$ and 
$M_s\sim {\cal O}(m_E)$.

{\em HTL one-loop pressure:}
Our starting expression is the one-loop HTL pressure, 
calculable from the graph in Fig.\ref{fig:ringtoHTL}. 
In the present cold quark matter
context it is evaluated at $T=0$, with (yet unspecified) mass $m_g$,
obtaining
\bea
\label{PHTL1loop}
&P_{\rm LO}^{\rm HTL}=\frac{d_A\ m_g^4}{(8\pi)^2} \left[ \frac{1}{2\epsilon}+ 
C_{11}-L +\epsilon \left(L^2+C_{21} L+C_{22}\right) \right] \nn \\
& \equiv m_g^4 \left[ -\frac{a_{1,0}}{2\epsilon} + a_{1,0} L +a_{1,1} +{\cal O}(\epsilon) \right]
\eea
where $L=\ln(\frac{m_g}{M})$ and $M$  is an arbitrary $\ms$
scale. 
In Eq.(\ref{PHTL1loop}) we also introduced a convenient notation for later purpose.
The coefficient $C_{11}\sim 1.17201$ was first obtained in \cite{HTLptT0}.
As explicited below, however, to work out complete NLL expressions we also need the ${\cal O}(\epsilon)$ terms, 
not previously available to our knowledge, that we have evaluated. 
We 
obtain $C_{21}= -2 C_{11},
C_{22}\simeq 2.16753$ (details can be found in 
\cite{suppmat}).\\ 
{\em RG resummation:}
We consider the renormalization group (RG) operator (see e.g. \cite{Collins}), that parametrizes the scale
variation in a (massive) theory:
\be
M \frac{d}{dM} \equiv  M \frac{\partial}{\partial M} + \beta(g^2)
  \frac{\partial}{\partial g^2} - \gamma^g_m(g^2) m_g
  \frac{\partial}{\partial m_g} \,,
\label{RG}
\ee 
where $M$ is a renormalization scale, $\beta(g^2)\equiv \mbox{d} g^2/\mbox{d}\ln M$ dictates the running coupling, and
the anomalous mass dimension $\gamma^g_m(g^2)\equiv \mbox{d} (\ln m_g)/\mbox{d}\ln M $ is further discussed below. 
In our convention with $g^2\equiv 4\pi \alpha_s$,
\be
\beta(g^2) = - 2b_0^g g^4-2b_1^g g^6+\cdots,\;
\gamma^g_m(g^2) = \gamma^g_0 g^2 +\gamma^g_1 g^4+\cdots
\label{betgam}
\ee
where $b_0^g,b_1^g$ are the QCD {\em pure gauge} contributions
\be
(4\pi)^2 \, b_0^g =\frac{11N_c}{3} , \;
(4\pi)^4 \, b_1^g = \frac{34 N_c^2}{3} .
\label{beta}
\ee
Given that HTL involves a gluon mass term $m_g$, 
from a RG standpoint\cite{rgopt_phi4} we consider $m_g$ blind to its precise dynamical origin 
as a screening mass, motivating to use the massive RG Eq.(\ref{RG}). 
Indeed, despite the nonlocal HTL Lagrangian, the NNLO perturbative HTL calculations 
give new $m_g$-dependent UV divergences and related counterterms having a
renormalizable form, as will be seen below, thus defining EFT anomalous dimensions in standard fashion.
It is important to stress
that the RG coefficients for such a massive theory are $T=0$ entities by definitions, even though they enter thermal or in-medium
contributions as well.
More precisely, within $T \ne 0$ HTL calculations, the divergences from $m_g \ne 0$ occur 
in two-loop order $\alpha_s (m^2_g T^2, m^3_g T)$ terms, and the corresponding
(unique) one-loop counterterm $\Delta m_g$ was obtained first in \cite{HTLpt2g}. 
Using this $\Delta m_g$ and the standard relation between bare mass $m_g^B$, $Z_{m_g}$ counterterm and Eq.(\ref{betgam}): 
$m^B_g \equiv m_g\,Z_{m_g} \simeq m_g (1-g^2\,\gamma_0^g/(2\epsilon) +{\cal O}(g^4))$, we easily identify:
\vspace{-2mm}
\be
\gamma^g_0 = \frac{11N_c}{3 (4\pi)^2} \equiv b_0^g.
\label{ga0b0}
\ee
Namely, the LO interactions from HTL that contribute to renormalize $m_g$ 
give a divergent contribution identical to the one defining $b_0^g$.
Although striking, this equality of pure gauge $b_0^g$ and $\gamma_0^g$ is 
merely a one-loop order accident. 
Incidentally, it is worth noting that the same result (\ref{ga0b0}) was obtained independently
from a localizable, {\em renormalizable} gauge-invariant setup for a (vacuum) gluon mass\cite{mgvac}: 
this is not a coincidence since those universal RG quantities are vacuum quantities independent of $T,\mu$.
The two-loop order $\gamma_1^g$, entering the NLO RG Eq.(\ref{betgam}) in our construction, has also  
been calculated from the same $T=\mu=0$ formalism, with the result\cite{mgvac2} $(4\pi)^4 \gamma_1^g = 77 N_c^2/12$. 
Furthermore, the mass renormalization within Eq.(\ref{PHTL1loop}), $P_{\rm LO}^{\rm HTL}(m_g\to m_g Z_{m_g} )$, 
generates additional terms that combine with genuine two-loop contributions Eq.(\ref{Psoftg3}).
Importantly,  
the unwanted nonlocal $\ln(m_E/M_s)/\epsilon$ divergence in Eq.(\ref{Psoftg3}) exactly cancels in those combinations,
while (local) remnant divergences after mass renormalization are renormalized by vacuum energy ${\cal E}_0$ ($\equiv -P$) 
counterterms, always necessary in a massive theory. 
According to Weinberg's theorem\cite{Collins}, such local counterterms 
prove the renormalizability of the $T=0$ HTL pressure 
at NLO $\alpha_s m_g^4$, i.e. NNNLO $\alpha_s^3$. 
The LO vacuum energy counterterm is the one
determined in HTL\cite{HTLpt2g,HTLptT0}, $\Delta {\cal E}_0^{(1)}= d_A m_g^4/(8\pi)^2/(2\epsilon)$.
At NLO its expression is given in 
\cite{suppmat}.
Remark that renormalizing $m_g$ in Eq.(\ref{PHTL1loop}) also modifies the {\em finite} coefficients in Eq.(\ref{p210}), 
as 
\bea
&2p_{-2} \rightarrow \! p_{-2},\;
p_{-1} \to \! p_{-1} -\frac{8\pi \ga0^g}{N_c} (C_{11}-\!\frac{1}{4}) \simeq p_{-1}\!\! -0.5381, \nonumber\\
&  p_0 \to  p_0 -\frac{8\pi \ga0^g}{N_c} (C_{22}-\frac{C_{11}}{2}) \simeq p_0 -0.9229 .
\label{p210mod}
\eea
In \cite{Gorda:2021kme}  
nontrivial cancellation mechanisms between soft $\alpha_s m_E^4$ 
and (presently not known) $\alpha_s^3$ hard contributions are 
convincingly argued to remove the divergent terms in Eq.(\ref{Psoftg3}), 
and to obtain the correct $\alpha_s^3 \ln^2 m_E$ coefficient, $2p_{-2}\to p_{-2}$.
In our massive renormalized scheme 
 the right $p_{-2}$ originates directly from standard mass renormalization cancellation mechanisms.
 This simple alternative picture appears consistent with generic EFT constructions\cite{EFTrev}, 
where all EFT UV divergences are renormalized, defining EFT anomalous dimensions, sufficient to extract $\ln m_E$
dependencies from the RG.\\
Having identified these key RG ingredients at $T=0$, obtaining
the LL and NLL at successive orders 
follows a well-established procedure\cite{Collins}.
Power counting\cite{Gorda:2021kme} dictates the soft pressure expansion to all orders
\be
P^{soft} \sim m^4_g \sum_{p=1}^\infty \left(g^2\right)^{p-1} \: \sum_{l=0}^p a_{p,l} \ln^{p-l}
(\frac{m_g}{M}) 
\label{Pgen}
\ee
with $M$ the ($\ms$ scheme) renormalization scale. The $a_{p,0} \ln^p (m_g/M), p\ge 1$ 
are the leading logarithmic (LL), $a_{p,1} \ln^{p-1}(m_g/M), p\ge 2$ the
next-to-leading logarithmic (NLL) coefficients and so on, with $a_{p,p}$ the non-logarithmic
coefficients at successive orders. 
Upon applying the RG Eq.(\ref{RG}) on Eq.(\ref{Pgen}) considering
$g$ fixed, 
we obtain recurrence relations. First for the LL series
\be
-p\: a_{p,0}= [4 \gamma_0^g +2b_0^g (p-2)] a_{p-1,0}, \;\;   p\ge 2
\label{LL}
\ee
where $a_{1,0}$ is given in Eq.(\ref{PHTL1loop}).
Before proceeding it is worth to give a first concrete outcome of Eq.(\ref{LL}),
that immediately gives at NLO:
\be
a_{2,0} = -2\gamma_0^g a_{1,0} \equiv -2b_0^g a_{1,0}
\label{LLNLO}
\ee
determining the LL term: $g^2\, m_g^4\, a_{2,0} \ln^2 (m_g/M)$, that upon using 
Eqs.(\ref{PHTL1loop}),(\ref{beta}) gives $(4\pi) a_{2,0} = N_c d_A/(8\pi)^2\, p_{-2}$,  
that perfectly matches Eqs.(\ref{Psoftg3}),(\ref{p210}) for $m_g\equiv m_E$ with
$p_{-2}$ in (\ref{p210mod}). 
 We stress that all LL in Eq.(\ref{LL}) rely on the {\em sole} one-loop $\ln (m_g/M)$ coefficient $a_{1,0}$ in Eq.(\ref{PHTL1loop}),  
thus are completely independent from the results in \cite{Gorda:2018gpy,Gorda:2021kme},
where $m_E^4 \alpha_s \ln^2 m_E$ was obtained from involved two-loop HTL calculations. The important feature
to obtain it from the above simple RG relations is having identified $\gamma_0^g$. 
More interestingly, Eq.(\ref{LL}) gives all higher order LL coefficients, a new result that we resum explicitly below.\\
One obtains similarly the NLL series (defined for $p\ge 2$)
\bea
&&(1-p) a_{p,1}= [4\gamma_0^g +2b_0^g (p-2)] a_{p-1,1} \nn \\
&& +  [4 \gamma_1^g +2b_1^g (p-3)] a_{p-2,0}  +\gamma_0^g (p-1)  a_{p-1,0}. \;\; 
\label{NLL} 
\eea
We recall that at a given perturbative order $g^{2p}$, the only new terms to calculate are
the single logarithm $a_{p,p-1}$ and nonlogarithmic $a_{p,p}$ terms. 
Both LL and NLL series above are convergent and can thus be resummed.\\
{\em Perturbatively RG invariant pressure:} 
Before giving resummed LL and NLL expressions, it is convenient to develop
a related important ingredient of our construction, namely to restore a RG invariant (RGI) massive pressure.
Indeed, applying Eq.(\ref{RG}) to Eq.(\ref{PHTL1loop}) gives a remnant scale dependence at leading HTL order, 
$\sim m_g^4 \ln M$, a well-known feature of massive theories. 
The appropriate way to deal with this is to realize that the RG Eq.(\ref{RG}) has an inhomogeneous term,
defining a 
(gluon) vacuum energy anomalous dimension $\hat\Gamma_0^g(g^2)=\Gamma_0^g +\Gamma_1^g g^2+\cdots$,
by analogy with other massive sector vacuum energies\cite{E0anomdim,E0phi4}:
\be 
\frac{d\,P^{\rm HTL}}{d\ln M}\equiv -m_g^4 \hat\Gamma_0^g(g^2)
=\frac{d}{d\ln M} (m_g^4 \sum_{k \ge 0} s_k^g g^{2k-2}) ,
\label{RGanom}
\ee
$\hat\Gamma_0^g(g^2)$ being
related to the vacuum energy counterterms (see 
\cite{suppmat}).
Thus an RGI combination is $P_{RGI}\equiv P^{\rm HTL}-m_g^4 \sum_{k} s_k g^{2k-2}$, where the $s_k^g$ coefficients 
are most simply determined\cite{rgopt_phi4} perturbatively from the second equality in Eq.(\ref{RGanom}):
at LO and NLO we obtain respectively
\be
s_0^g = -\frac{a_{1,0}}{2(b_0^g-2\gamma_0^g)} = \frac{-d_A}{2(8\pi)^2\,b_0^g},
\label{s0g}
\ee
\be
s_1^g = a_{1,1}+ \frac{a_{2,1}}{4\gamma_0^g} +\frac{a_{1,0}}{4} +
\frac{s_0^g}{2\gamma_0^g}(b_1^g-2\gamma_1^g),
\label{a11def}
\ee
where $a_{1,1}= -a_{1,0} C_{11}$ in Eq.(\ref{PHTL1loop}). 
Embedding Eq.(\ref{s0g}) within Eq.(\ref{LL})\footnote{Since $s_0^g, s_1^g$ are obtained
respectively from $a_{1,0}, a_{2,1}$
by applying the RG Eq.(\ref{RG}), we conveniently 
recast the LL, NLL series such that $s_0^g$ defines their first term, i.e. 
with $p\ge 1$ in Eq.(\ref{LL}) and $a_{0,0}\equiv -s_0^g$.} one can easily resum the LL series as
\bea
&P^{\rm sum}_{\rm LL}=-\frac{s_0^g\:m_g^4}{g^2} f_1^{1-4(\frac{\gamma_0^g}{2b_0^g})} =-\frac{s_0^g \:m_g^4}{g^2} f_1^{-1}, \nn \\
& f_1 = 1+2b_0^g\,g^2 \ln \frac{m_g}{M},
\label{LLsum}
\eea
where 
we also used Eq.(\ref{ga0b0}).
It is straightforward to check that Eq.(\ref{LLsum}) reproduces at all orders the coefficients in Eq.(\ref{LL}).
The last equality in Eq.(\ref{LLsum}) tells that the LL series iterates simply like the (pure gauge) running coupling,
but this is merely an accident of LO RG Eq.(\ref{ga0b0}).\\
Similarly, after more algebra one can resum formally the NLL series.
 Adapting results from\cite{jlrgsum} we obtain:
\be
P^{\rm sum}_{\rm NLL}= \frac{-s_0^g \,m_g^4}{g^2\,f_2^{4 A_0-1}}
[R(f_2)]^B 
(1 -\frac{a^\prime_{1,1}\,g^2}{s_0^g\, f_2} 
-\frac{a_{2,2}\,g^4}{s_0^g\,f_2^2}),
\label{NLLsum}
\ee
\bea
\ds
&R(f_2)= (1 + g^2\, b_1^g/(b_0^g f_2) )/(1 +  g^2 \,b_1^g/b_0^g ), \nn \\
&\!\!\! f_2= 1+\left[ 2b_0^g\, g^2 +2(b_1^g -\gamma_0^g b_0^g)g^4\right] \ln \frac{m_g}{M} +{\cal O}(g^6),
\label{f2def}
\eea
and 
$A_0= \gamma_0^g/(2b_0^g), A_1= \gamma_1^g/(2b_1^g), B= 4(A_1-A_0)$.  
The exact $f_2$ expression, reproducing Eq.(\ref{NLL}) to all orders from Eq.(\ref{NLLsum}),
is given in 
\cite{suppmat}.
Eq.(\ref{f2def}) gives numerically good approximations 
as long as the coupling is not too large 
($\alpha_s \lesssim 0.5$).\\
A few remarks are worth regarding the input content of Eq.(\ref{NLLsum}):
i) It is rather generic, 
but numerically relies on the lowest order 
{\em purely soft} NLL coefficient $a_{2,1}\propto p_{-1}$ in Eq.(\ref{Psoftg3}), calculated in \cite{Gorda:2021kme}.  
 Accordingly, Eq.(\ref{NLLsum}) does not include the (presently unknown) QCD {\em mixed} soft-hard NLL 
contributions, see \cite{Gorda:2021kme}.
ii) One obtains $a'_{1,1} \equiv a_{1,1}-s_1^g$ 
with $a_{1,1}$ defined in Eq.(\ref{PHTL1loop}), 
since the NLO subtraction coefficient $s_1^g$ in Eq.(\ref{a11def}) 
contributes a correction to $a_{1,1}$.
iii) $a_{2,2}$ in Eq.(\ref{NLLsum}) incorporates the ${\cal O}(g^2 m_g^4)$ non-logarithmic term
$p_0$ in Eq.(\ref{Psoftg3}): the
precise connection between the parameters in Eq.(\ref{NLLsum}) and $p_{-1},p_0$ in Eqs.(\ref{Psoftg3}),(\ref{p210mod}) 
is $a_{2,1}= N_c/(4\pi) a_{1,0} (2p_{-1})$, $a_{2,2} = -N_c/(4\pi) a_{1,0}\, p_{0}$, 
{\em after} modifying $p_{-1}, p_0$ in Eq.(\ref{p210mod}).\\
{ \em Soft and hard pressure matching:} 
The massive RG construction above basically concerns the soft pressure contributions,
thus $\ln m_g/M \to \ln m_E/M_s$,
with overall factor $m_E^4\sim \alpha_s^2$, see Eq.(\ref{mEdef}). 
While the hard contribution in Eq.(\ref{Pas2}), known exactly only at $\alpha_s^2$-order, 
is added perturbatively to Eqs.(\ref{LLsum}),(\ref{NLLsum}). 
The latter RGI expressions formally cancel
the soft scale dependence up to neglected ${\cal O}(g^2 m_E^4)$,  ${\cal O}(g^4 m_E^4)$ terms respectively.
At ${\cal O}(\alpha_s^2)$,
it is equivalent to the complete soft scale cancellation operating in the factorization picture, thus, we can choose any $M_s$.
While at $\alpha_s^{p\ge 3}$-orders, partial ignorance of hard and mixed contributions incites to take 
$M_s\sim {\cal O}(m_E)$ in the $\ln m_E/M_s$ terms.\\
Another feature to account while combining the 
soft and hard contributions is that Eq.(\ref{NLLsum}) entails
a nonlogarithmic term $\sim \alpha_s^2 a_{1,1}$ with $a_{1,1}$ from Eq.(\ref{a11def}), obviously different from the genuine 
``soft +hard'' $\alpha_s^2$ term in Eq.(\ref{Pas2}). Since only soft contributions are RG-resummed, 
to avoid wrong contaminations in nonlogarithmic hard terms 
from RG-induced NLL soft terms, one should perturbatively subtract $a_{1,1} m_E^4$ from the total expression 
(which does not affect higher order NLL terms generated by $a_{1,1}$ within Eq.(\ref{NLLsum})).\\
 As a last important subtlety, note that although the subtraction terms
as above determined, Eqs.(\ref{s0g}),(\ref{a11def}), are sufficient to define LO and NLO RGI pressures 
Eqs.(\ref{LLsum}),(\ref{NLLsum}), 
actually the complete integration of Eq.(\ref{RGanom}) entails an extra boundary condition 
(see 
\cite{suppmat}):
\be P^{\rm BC}_{\rm RGI}=  m_g^4(M_0)(\frac{s_0^g}{g^2(M_0)}+s_1^g),
\label{PRGIBC}
\ee
of similar form as the subtraction terms but involving a (boundary) scale $M_0\ne M$.
We
can use this freedom to 
set $M_0$ such that Eq.(\ref{PRGIBC}) provides 
an appropriate EFT-matching\cite{EFTrev} of the soft pressure to the full one at ${\cal O}(\alpha_s^2)$,
i.e. with $M_0\sim {\cal O}(\mu)$,
consistently also with the Stefan-Boltzmann limit.\\
Collecting all, our final pressure expression 
is obtained upon formally replacing $P^{\rm soft}_{\alpha_s^3}$ in Eq.(\ref{Pas2}) 
by
\be
P^{\rm sum}_{\alpha_s^{p\ge 3}} = P^{\rm sum}_{\rm (N)LL} 
 +P^{\rm BC}_{\rm RGI}  
-m_E^4 (a_{1,1} +a_{1,0}\ln \frac{m_E}{M_s}) 
-P^{\rm match}_{\alpha_s^2}
\label{Pfin}
\ee
with $a_{1,1}=0$, $s_1^g\equiv 0$ for $P^{\rm sum}_{\rm LL}$, and 
$P^{\rm match}_{\alpha_s^2}$ given in Eq.(B10) in \cite{suppmat}.
The last three terms in Eq.(\ref{Pfin}) are required to match Eq.(\ref{Pas2}) consistently 
(i.e. without double counting), so that all RG-induced extra terms are ${\cal O}(\alpha_s^3)$.\\

\begin{figure}[!]
\centerline{ \epsfig{file=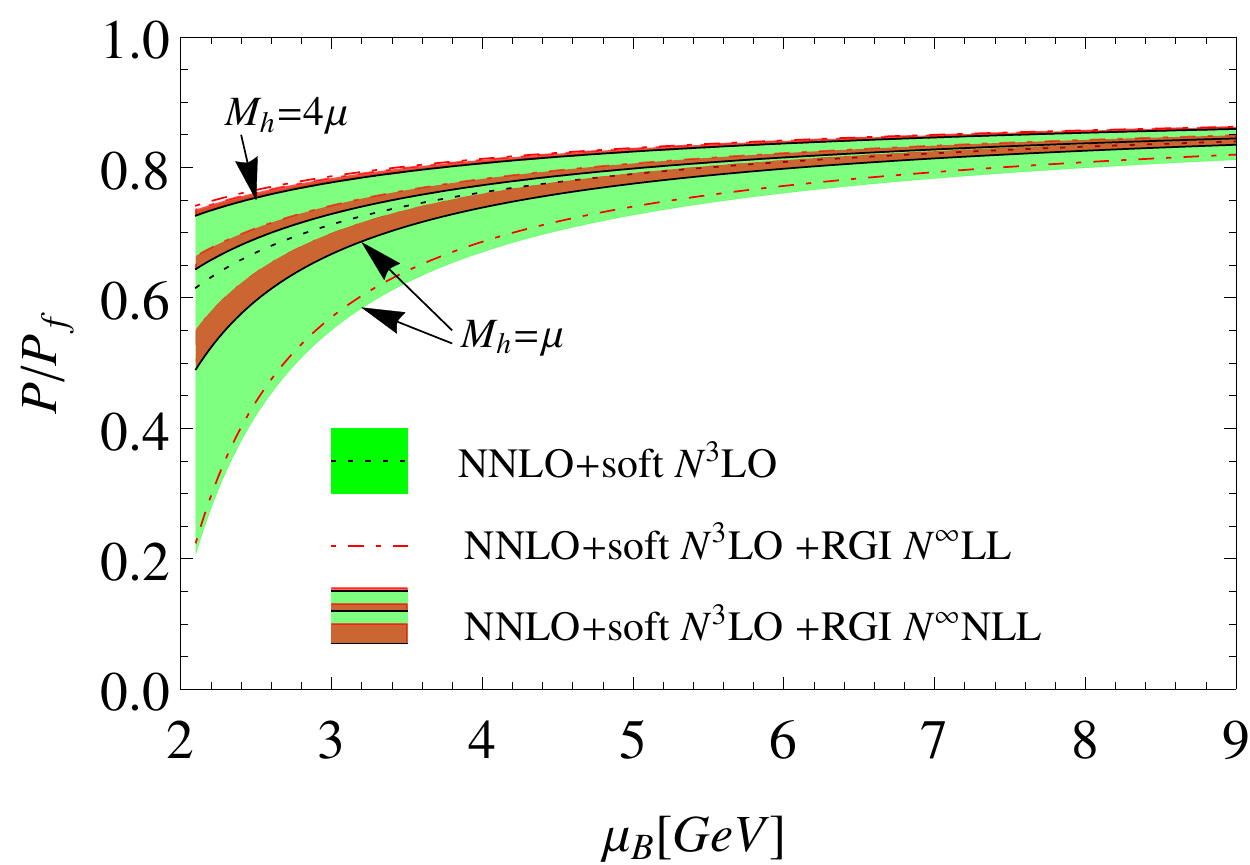,width=1.\linewidth,angle=0}}
\caption{NNLO + soft $\rm N^3LO$ Eqs.(\ref{Pas2}),(\ref{Psoftg3})
versus NNLO + soft $\rm N^3LO$ +RGI LL, NLL resummed 
pressures, as function of $\mu_B=3\mu$, with $\mu \le M_h\le 4\mu, M_s\simeq 0.275 m_E$, $M_0=2\mu$.
 For the RGI NLL resummed pressure, $M_s$ variations within $[M_s/2, 2M_s]$ are shown in addition as darker bands. }
\label{figPbands}
\end{figure}

{\em Numerical results and comparisons:}
In Fig.\ref{figPbands} the RGI LL and NLL resummed pressures from Eq.(\ref{Pfin})\footnote{See also 
Eqs.(B11),(C6) in \cite{suppmat} for compact expressions.}
are compared to the present state-of-the-art Eqs.(\ref{Pas2}),(\ref{Psoftg3})
as function of $\mu_B=3\mu$. The central scale values $M_h=2\mu$ and the $\mu \le M_h \le 4\mu$ 
remnant scale dependence are illustrated for the different quantities,
using in Eqs.(\ref{Pas2}),(\ref{Pfin}) the exact NLO QCD running coupling $\alpha_s(M_h)$ 
with\cite{PDG2018} $\Lambda_{\ms} \sim 0.32$ GeV.
For sensible comparisons we also adopt the minimal sensitivity-determined\cite{pms} 
soft scale in \cite{Gorda:2021znl}, $M_s\sim 0.275 m_E$. 
 Finally, we fix $M_0=2\mu$ in Eq.(\ref{PRGIBC}), a natural choice as it calibrates 
Eq.(\ref{Pfin}) to the central $M_h$ values of the NNLO pressure.\\
Note first importantly that the sole LL resummation,  
given by Eq.(\ref{LLsum}) with $-s_0^g\to a_{1,0} g^2 \ln m_g/M$,
gives a sizeably reduced scale dependence, compared to the NNLO pressure
Eq.(\ref{Pas2}) at ${\cal O}(\alpha_s^2)$, as
Eq.(\ref{LLsum}) induces {\em positive} $\alpha_s^{p\ge 3}$ contributions partly
cancelling the negative $\alpha_s^2$ coefficient in Eq.(\ref{Pas2}).
However, this effect is approximately cancelled once including $p_{-1},p_0$ $\alpha_s^3$-order terms 
Eqs.(\ref{p210}),(\ref{p210mod}).
Next, for the LL and NLL RGI pressures, deviations from the state-of-the-art
(``NNLO + soft $\rm N^3LO$'' in Fig.\ref{figPbands}) are noticeable. 
The central scale ($M_h=2\mu$) RGI pressure is slightly 
higher for fixed $\mu$ values, 
with very moderate differences between LL and NLL pressures. 
Importantly, the remnant scale dependencies of the resummed pressures are reduced
as compared to NNLO + soft $\rm N^3LO$ results: only slightly for the LL pressure, due to 
cancellations with $p_{-1},p_0$ ${\cal O}(\alpha_s^3)$ terms, 
but significantly for the NLL one,
both for $M_h$ and $M_s$ variations\footnote{Note that $M_s\sim 0.275 m_E$ {\em maximizes} the $M_h$ variations. 
As we also checked, uncertainties from varying $M_0$, for fixed $M_h$, are much smaller than the $M_h$ variations 
in Fig.\ref{figPbands}.}, due to 
$\alpha_s^{p\ge 3}$ terms induced both by NLL 
and the RGI-restoring terms 
in Eqs.(\ref{NLLsum}),(\ref{Pfin}). 

In conclusion, we have obtained compact explicit expressions for the all order ``double'' resummations of
LL and NLL soft contributions to the cold and dense QCD pressure, that goes well beyond 
previously established results. Our
RG resummation construction moreover gives clearly improved residual scale dependence. 
This should provide improved control towards 
lower $\mu_B$ values to match with the extrapolated EoS from the nuclear matter density region. 
We thus anticipate that our present results may have strong implications once embedded within a more realistic EoS. 
It is not difficult to extend our framework to include different chemical potential and nonzero masses for the quarks, 
in order to more realistically describe the EoS relevant for beta equilibrium and neutron star properties,
that we leave for future investigation.\\

{\em Acknowledgments}: 
We thank Marcus B. Pinto and Aleksi Vuorinen for discussions. \\
\newpage
\onecolumngrid

\appendix
\section{One-loop $T=0$ HTL pressure calculation}\label{HTLlo}
We give here the main steps 
of our derivation of Eq.(\ref{PHTL1loop}), where the ${\cal O}(\epsilon)$ contributions 
is a new result, to the best of our knowledge. For the HTL formalism and relevant expressions 
of the gluon propagator and self-energies we refer e.g. to \cite{HTLpt2g,HTLrev2020}. 
We mainly follow the convenient Euclidean $T=0$ conventions and expressions in \cite{Gorda:2021kme}, to which we
refer for more details here skipped.
The one-loop graph in the right of Fig.1
gives for the HTL free energy (see e.g. \cite{HTLpt2g,HTLptT0}): 
\be
\Omega^{\rm HTL}_{\rm LO} = -\frac{d_A}{2} \int_K \left( (d-1) \ln [\Delta_T(K)] +\ln [\Delta_L(K)] \right), 
\label{OmLO}
\ee  
where $K^\mu= (K_0, {\bf k})$ is an Euclidean vector in $d+1$ dimensions (the integration measure 
in (\ref{OmLO}) will be specified below), 
\begin{equation}\label{eq:Gi_gluons}
\Delta_{T,L}(K)\equiv \frac{1}{K^2+\Pi_{T,L}(K)} ,
\end{equation}
$\Pi_{T,L}$ being the transverse and longitudinal components of the 
one-loop HTL self-energy tensor
\begin{equation}\label{eq:HTLselfEnergyTensor}
\Pi^{\mu\nu}(K)= T^{\mu\nu}(\hat{K})\Pi_T(K)+L^{\mu\nu}(\hat{K})\Pi_L(K),
\end{equation}
where the $T,L$ projection operators are
\be
\label{eq:DdimProjecteurOperator}
\begin{aligned}
T^{\mu\nu}(\hat{K})& \equiv \delta^{\mu i}\delta^{\nu j}\left(\delta^{ij}-\hat{k}^i\hat{k}^j\right)\\
L^{\mu\nu}(\hat{K})& \equiv \delta^{\mu\nu}-\hat{K}^{\mu}\hat{K}^{\nu}-T^{\mu\nu}(\hat{K}) \\
\end{aligned}
\end{equation}
with $\hat{K}= K/|K|$ and $\hat{\bf k}=\bf k/|\bf k|$.\\
For the HTL approximation relevant for cold quark matter, 
one has
\begin{equation}\label{eq:SelfEnergy}
\Pi^{\mu\nu}(K)=m_g^2 \int_{\hat{\mathbf{v}}}\left(\delta^{\mu 0}\delta^{\nu 0}-\frac{i K_0}{K\cdot V}V^{\mu}V^{\nu}\right)
\end{equation}
where $V^{\mu}\equiv (-i,\hat{\mathbf{v}})$ is a lightlike vector,
with $\hat{\mathbf{v}}$ a $d$-dimensional unit vector. 
Taking the trace and $00$ components of Eq.(\ref{eq:HTLselfEnergyTensor}), 
and Eq.(\ref{eq:SelfEnergy})
with appropriate $d$-dimensional measures, leads to 
\bea
& \Pi^{\mu\mu}(K)=m_g^2 \int_{\hat{\mathbf{v}}}\delta^{00} =m_g^2\\
& \Pi^{00}(K)= m_g^2 \left[1+ \int_{\hat{\mathbf{v}}}\frac{i K_0}{-i K_0+|{\mathbf{k}}|z_v}\right]\nn \\
& =m_g^2 \left[1-_{2}\!{F}_1\left(\frac{1}{2},1,\frac{d}{2};-\frac{{\mathbf{k}}^2}{K_0^2}\right) \right],
\label{Pi002F1}
\eea
where $_{2}\!\,{F}_1$ is the hypergeometric function.
From Eq.(\ref{Pi002F1}) it is not difficult to derive the Taylor expansion in $\epsilon$ of $\Pi^{00}$, thus of
$\Pi_L, \Pi_T$
from successive derivatives $\partial_c\,[ _{2}\!{F}_1(a,b,c;z)]$.
Finally, expressing $\Pi_{L,T}$ in terms of the Euclidean $\phi_K$ angle, $\tan\phi_K\equiv |{\bf k}|/K_0$,
gives 
\bea
&&\Pi_{T}(\phi_K)= \frac{m_g^2}{2}\cot(\phi_K)\left[\bar \phi_K \csc^2(\phi_K)-\cot(\phi_K)\right]+{\cal O}(\epsilon)\nn \\
&&\Pi_{L} (\phi_k)= m_g^2 \csc^2(\phi_K)\left[1-\bar \phi_K\cot(\phi_K)\right] +{\cal O}(\epsilon),
\label{eq:PiEps0Simplifie}
\eea
where 
$\bar \phi_K\equiv \arctan[\tan(\phi_K)]$,
and similarly for higher order terms in $\epsilon$.
From such expressions we can evaluate Eq.(\ref{OmLO}), 
with $\int_K \equiv 2\pi^{d/2}/\Gamma(d/2)/(2\pi)^{(d+1)}\int_0^\pi d \phi_k \sin^{d-1} (\phi_k) 
\times \int_0^\infty dk\, k^d $.
The integral over $k\equiv |K|$ is easily performed analytically, with the leading UV-divergent term $\sim 1/\epsilon$ 
extracted. 
To obtain next the finite and ${\cal O}(\epsilon)$ coefficients requires to
expand $\Pi_{L,T}$ up to order $\epsilon^2$ in Eq.(\ref{eq:PiEps0Simplifie}). 
Upon numerical remnant angular integration we obtain after algebra the $T=0$
pressure $P_{\rm LO}^{\rm HTL} = -\Omega_{\rm LO}^{\rm HTL}$ in the $\ms$-scheme:
\bea
&P_{\rm LO}^{\rm HTL}= -
 (\frac{e^{\gamma_E} M^2}{4\pi})^\epsilon\,
\frac{d_A}{2-\epsilon} \,\sec(\pi \frac{d}{2})\frac{2\pi^{\frac{d}{2}}}{\Gamma(\frac{d}{2})} \nn \\
& \times (\frac{m_g}{2\pi})^{d+1} (\frac{\pi^2}{16})
\left(1+ d_{11}\epsilon + d_{12}\epsilon^2 \right)
\label{PHTLbare}
\eea
where $M$ is the $\ms$ arbitrary scale and
\be 
d_{11} \simeq 1.23032,\;
d_{12} \simeq 1.04214,
\ee
that upon expanding Eq.(\ref{PHTLbare}) in $\epsilon$ with $d=3-2\epsilon$ gives Eq.(\ref{PHTL1loop})
with 
\bea
&&C_{11} = -\frac{1}{2} C_{21}= \frac{5}{4}-\ln 2 +\frac{d_{11}}{2}\simeq 1.17201, \nn \\
&& C_{22}= (\frac{5}{4}-\ln 2) d_{11} +\frac{d_{12}}{2}+\ln^2 2-\frac{5}{2}\ln 2+\frac{21}{8}-\frac{\pi^2}{24} \simeq 2.16753,
\eea
where $C_{11}$ was first obtained in \cite{HTLptT0}.
\section{Vacuum energy anomalous dimension and RG-invariant pressure}
Once combining the genuine two-loop contributions Eq.(\ref{Psoftg3})
and Eq.(\ref{PHTLbare}) with renormalized mass, $P_{\rm LO}^{\rm HTL}(m_g\to m_g Z_{m_g} )$ where
$Z_{m_g} \simeq (1-g^2\,\gamma_0^g/(2\epsilon) +{\cal O}(g^4))$ in our normalization,
the $\ln (m_E)/\epsilon$ in Eq.(\ref{Psoftg3}) cancels, and remnant local divergences are 
renormalized by minimal subtraction, defining the NLO vacuum energy counterterm: 
\be
\Delta {\cal E}_0^{(2)} = \frac{d_A\,m_g^4}{(8\pi)^2} \left\{ \frac{1}{2\epsilon}+ 
N_c (\frac{g^2}{4\pi}) \left( -\frac{11}{24\pi\,\epsilon^2}  
 + [\frac{p_{-1}}{2\epsilon}-\frac{11}{6\pi\epsilon}(C_{11}-\frac{1}{4})] \right) \right\} \equiv Z_0^g(g^2)m_g^4,
\label{vacEn2}
\ee
where the LO term was derived in standard HTL\cite{HTLpt2g}. 
As remarked in the main text, the resulting finite pressure is not RG invariant: rather,
Eq.(\ref{vacEn2}) implies a gluon vacuum energy density anomalous dimension $\hat\Gamma_0^g(g^2)$, 
modifying the homogeneous RG equation as:
\be
\frac{d\,{\cal E}_0}{d\ln M} [=-\frac{d\,P^{\rm HTL}}{d\ln M}] \equiv \hat\Gamma_0^g(g^2) m^4_g,  
\label{RGanomSM}
\ee
where $\hat\Gamma_0^g(g^2)=\Gamma_0^g +\Gamma_1^g g^2+\cdots$.
In other words, the massive pressure (equivalently vacuum energy), considered as a dimension four EFT operator, 
starts to mix with the unit operator $\sim m_g^4 \mathds{1} $ already at one-loop. 
$\hat\Gamma_0^g(g^2)$ can be obtained from the counterterm Eq.(\ref{vacEn2}):
we define (see e.g. \cite{E0anomdim})
\be
{\cal E}_0^B \equiv M^{-2\epsilon} \left( {\cal E}_0(g^2)-m_g^4\,Z_0^g(g^2) \right)
\ee
with ${\cal E}_0^B, {\cal E}_0$  the bare and renormalized vacuum energies respectively and $Z_0^g$ the (dimensionless) counterterm 
in Eq.(\ref{vacEn2}). Then applying the RG equation (\ref{RG}): $d({\cal E}_0^B)/d\ln M \equiv 0$ (with $\beta(g^2)\to 
\overline \beta(g^2)\equiv -2\epsilon g^2+\beta(g^2)$ as it is appropriate for bare quantities), 
 we obtain after straightforward algebra
\be
\hat\Gamma_0^g(g^2) = 
(-4\gamma_m(g^2) -2\epsilon) Z_0^g(g^2) +\overline\beta(g^2) \frac{\partial}{\partial g^2} Z_0^g(g^2).
\label{vacen}
\ee
Up to NLO this gives
\be
\Gamma_0^g= a_{10} = -\frac{d_A}{(8\pi)^2},\;\; 
\Gamma_1^g = 2 a_{10} \left(\frac{N_c}{4\pi}\right)\left(p_{-1}-\frac{11}{3\pi}(C_{11}-\frac{1}{4})\right) .
\ee

Next we consider 
the complete integral solution of the RG Eq.(\ref{RGanomSM}), formally obtained as (see e.g. \cite{Collins}) 
\be
P^{\rm HTL}(g^2(M),m(M))=P^{\rm HTL}(g^2(M_0),m(M_0))+ \int_{g^2(M_0)}^{g^2(M)} d x \left \{ \frac{- \hat\Gamma_0^g(x)}{\beta(x)} 
\exp \left[ -4 \int_{g^2(M_0)}^{x} d y \frac{\gamma_m(y)}{\beta(y)} \right] \right \} m^4_g(M_0) , 
\label{RGint}
\ee
where $M_0$ is a reference or ``initial'' scale.
Working out Eq.(\ref{RGint}) explicitly at perturbative NLO, using $\beta(g^2)$, $\gamma_m(g^2)$ from 
Eq.(\ref{betgam}), 
we obtain after some algebra the (NLO) RG-invariant combination 
\be
P^{\rm HTL}_{RGI} \equiv P^{\rm HTL} -\left[ m_g^4(M) \left(\frac{s_0^g}{g^2(M)}+s_1^g\right) 
-m_g^4(M_0)\left(\frac{s_0^g}{g^2(M_0)}  +s_1^g\right) +{\cal O}(m_g^4 g^2) \right].
\label{subtot}
\ee
Note that to obtain the final form of Eq.(\ref{subtot}) we have identified the NLO running mass expression:
\be
m_g(M) \simeq m_g(M_0) \left(\frac{g^2(M)}{g^2(M_0)}\right)^{\frac{\gamma_0^g}{2b_0^g}} 
\left[1+\frac{2}{(b_0^g)^2}(b_0^g \gamma_1^g -b_1^g\gamma_0^g)\,g^2(M_0)+{\cal O}(g^4)\right], 
\ee
and used the relations between $s_0^g, s_1^g$ in Eqs.(\ref{s0g}),(\ref{a11def}) and vacuum energy anomalous 
dimension coefficients $\Gamma_i^g$:
\be
\Gamma_0^g = -2 s_0^g (b_0^g-2\gamma_0^g); \;\;
\Gamma_1^g = 4\gamma_0^g s_1^g-2 s_0^g (b_1^g-2\gamma_1^g).
\ee
Note that all the terms $\propto m_g^4(M)$ in Eq.(\ref{subtot}), sufficient for restoring RGI with respect
to $M$, can be obtained in a more pedestrian way by working out perturbatively the second equality in 
Eq.(\ref{RGanom}), however 
missing the boundary terms $\propto m_g^4(M_0)$ of the complete solution Eq.(\ref{subtot}).
For sufficiently large $M_0$, the latter boundary terms 
behave as
$s_0^g [2b_0^g \ln(M_0/\Lambda_{\ms})]^{1-4\gamma_0^g/(2b_0^g)} \sim \ln^{-1}(M_0/\Lambda_{\ms})$, 
where $\Lambda_{\ms}$ is the basic QCD scale. The occurrence of $1/g^2$ terms in Eq.(\ref{subtot}) is  
related to the LO $\hat\Gamma_0^g(g^2)$ being ${\cal O}(1)$ in $g$, 
a consequence of the ${\cal O}(1)$ LO pressure $1/\epsilon$ divergence. 
However, the difference of the two terms in brackets in Eq.(\ref{subtot}) has its  
leading perturbative contribution starting at ${\cal O}(g^4)$, 
more precisely:
\be
P^{extra}_{\alpha_s^2} = P_f (\frac{2N_f}{\pi^2}) \alpha_s^2(M_0) (\frac{b_0}{b_0^g}) \ln \frac{M_0}{M} \equiv P^{match}_{\alpha_s^2}, 
\label{Pmatch}
\ee
where $b_0 \equiv b_0^g -(2/3) N_f/(4\pi)^2$, i.e. $b_0/b_0^g =9/11$ for $N_f=3$, accounts for the full QCD running
coupling used for the total (soft + hard) pressure.
We thus subtract Eq.(\ref{Pmatch}) in Eq.(\ref{Pfin}) to consistently match the NNLO pressure Eq.(\ref{Pas2})
up to higher order $\alpha_s^{p\ge 3}$ terms.\\
Collecting all relevant terms, a convenient compact expression for the total pressure with
Eq.(22) restricted to LL terms\footnote{ Remark that in Fig. 2 the ``NNLO + soft $N^3LO$ +RGI LL'' results 
accordingly include in addition to Eq.(\ref{PLLcompact}) the ${\cal O}(\alpha_s^3)$ soft contributions 
$\propto p_{-1}, p_0$ in Eq.(\ref{Psoftg3}), with $p_0, p_{-1}$ redefined in Eq.(\ref{p210mod}).} reads explicitly:
\be
P^{\rm sum,LL}_{\alpha_s^{p\ge 3}} =  P^{cqm}_{\alpha_s^2} + P_f (\frac{2N_f}{\pi^2}) 
\left(\frac{1}{8\pi\,b_0^g} \left[\frac{\alpha_s}{1+8\pi b_0^g\, \alpha_s \ln \frac{m_E}{M_s}} -\alpha_s(M_0)\right] +
\alpha_s^2 \ln \frac{m_E}{M_s} -\alpha_s^2(M_0) \frac{b_0}{b_0^g}\, \ln \frac{M_0}{M_h}\right),
\label{PLLcompact}
\ee
with $P^{cqm}_{\alpha_s^2} $ given in Eq.(\ref{Pas2}), $b_0^g$ in Eq.(\ref{beta}), and $\alpha_s\equiv \alpha_s(M_h)$. 
%
\section{Compact NLL resummation}\label{NLLcomp}
In this appendix we give some more details on the NLL resummation. 
Rather than working out the result Eq.(\ref{NLLsum}) directly from Eq.(\ref{NLL}) which is tedious, 
it is more convenient to equivalently first derive an exact NLO running mass expression 
$m_R(M^\prime) = m_R(M) \exp( -\int^{g^2(M^\prime)}_{g^2(M)} \gamma_m(g)/\beta(g))$ using Eq.(\ref{RG}),
with the boundary condition\cite{jlrgsum} $ m_R(m_R)\equiv m_R$.
Then, considering dimensionally dictated expression $\propto m_R^4$ appropriate to 
match a vacuum energy and adding nonlogarithmic perturbative contributions, 
it leads after some algebra to Eq.(\ref{NLLsum}), with 
\bea
\ds
& f_2 = 1+ 2 b_0^g\, g^2 \left \{ \ln \frac{m_g}{M}
     + (C-A_0) \ln f_2  \right. \nn \\
&  \left.    + (C+A_1-A_0) \ln R(f_2) \right \} \nn \\
&= 1+\left[ 2b_0^g\, g^2 +2(b_1^g -\gamma_0^g b_0^g)g^4\right] \ln \frac{m_g}{M} +{\cal O}(g^6).
\label{f2defapp}
\eea
where $A_0= \gamma_0^g/(2b_0^g)$, $A_1= \gamma_1^g/(2b_1^g)$,
$C= b_1^g/(2(b^g_0)^2)$ and 
$R(f_2)$ is defined after Eqs.(\ref{NLLsum}),(\ref{f2def}).
Notice that $f_2$ in Eq.(\ref{f2defapp}) is an implicit function, that should be iterated 
to correctly reproduce (analytically) the NLL coefficients in Eq.(\ref{NLL}) to all orders.
More conveniently truncating it at order $g^4$, i.e. using the last line of Eq.(\ref{f2defapp}),
numerically gives a very good approximation as long as the coupling is not too large.

Next, we give an alternative exact compact expression  
of Eq.(\ref{NLLsum}) in terms of explicitly RGI quantities, more convenient than 
the implicit relation in first line of Eq.(\ref{f2defapp}).
Defining the two-loop order RGI mass\cite{jlrgsum}
\be
\hat m_g= 2^C m_g (2b_0^g\,g^2)^{-A_0}(1+b_1^g\,g^2/b_0^g)^{A_0-A_1},
\label{mghat}
\ee 
 $F_2=f_2/(2b_0^g\,g^2)$,
and using 
the exact two-loop running coupling, 
implicit $g^2(M)$ solution of
\be
\Lambda_{\ms} = M e^{-1/(2b_0^g g^2)} (b_0^g\,g^2/(1+b_1^g\,g^2/b_0^g))^{-C} ,
\label{Lam2}
\ee
Eq.(\ref{NLLsum}) with $m_g\to m_E$ can be rewritten
\bea
&& P^{\rm sum}_{NLL}= -2b_0^g s_0^g 2^{-4C} \hat m_E^4 \, F_2^{1-4 A_1} (C+F_2)^{4(A_1-A_0)} \nn \\
&& \times \left(1 -\frac{a^\prime_{1,1}}{2s_0^g b_0^g F_2} -\frac{a_{2,2}}{s_0^g\,(2b_0^g F_2)^2}\right),
\label{NLLsumF}
\eea
where  $a^\prime_{1,1}= a_{1,1}-s_1^g$ and $F_2$ is the solution of
\be
e^{F_2} F_2^{A_1} (C+F_2)^{A_0-A_1-C}= \frac{\hat m_E}{\Lambda_{\ms}}
\label{F2num}
\ee
easily determined numerically for given $g^2=4\pi\alpha_S$, $m_E$ in Eq.(\ref{mEdef}) and $A_0, A_1, C$ coefficients given 
above. Explicitly, in Eqs.(\ref{mghat})-(\ref{F2num})
one has $b_0^g s_0^g = -1/(4\pi)^2$,
$A_0=1/2$, $A_1= 77/272$, $C=51/121$,  $a^\prime_{1,1}\simeq 0.02385$, $a_{2,2}\simeq 0.00390$.
The numerical difference between the exact expression Eq.(\ref{NLLsumF}) and the ${\cal O}(g^4)$ truncation in Eq.(\ref{f2def})
is smaller than $10^{-3}$ for $\alpha_S\le 0.5$ and $\le 10^{-2}$ for $0.5< \alpha_S\lesssim 1$. Once embedding Eq.(\ref{NLLsumF}) 
within the complete pressure the difference is hardly visible since the hard contributions largely dominate for small $\mu$ values.\\
Finally, similarly to Eq.(\ref{PLLcompact}), we provide a compact expression for the total pressure with RGI NLL-resummed 
contributions in Eq.(\ref{Pfin}):
\bea
 P^{\rm sum,NLL}_{\alpha_s^{p\ge 3}} &=&  P^{cqm}_{\alpha_s^2} + P_f (\frac{2N_f}{\pi^2}) 
\Bigg(\frac{1}{8\pi\,b_0^g} \Bigg[\frac{\alpha_s}{f_2} 
\left(\frac{1+\frac{51}{22\pi}\frac{\alpha_s}{f_2}}{1+\frac{51}{22\pi}\alpha_s}\right)^{-59/68}\!\!
\left(1+d_1\frac{\alpha_s}{f_2}+d_2\frac{\alpha_s^2}{f_2^2}\right)  -\alpha_s(M_0)\Bigg]  \nn \\
&&  + \alpha_s^2 \left(\ln \frac{m_E}{M_s}+d_3 \right) +d_4 \alpha_s^2(M_0) 
-\alpha_s^2(M_0) \frac{b_0}{b_0^g}\, \ln \frac{M_0}{M_h}\Bigg),
\label{PNLLcompact}
\eea
where\footnote{Note that $d_1$, $d_4$ 
depend on  $p_{-1}$ and $d_2$ on $p_0$ from Eq.(\ref{p210mod}),
thus related to the coefficients originally calculated in [18], see also Eq.(\ref{a11def}).} 
$d_1\simeq 3.29659 \equiv -(4 \pi/s_0^g) (a_{1,1}-s_1^g) $, 
$d_2\simeq 6.77276 \equiv -((4\pi)^2/s_0^g) a_{2,2}$,
$d_3\simeq -1.17201\equiv -8\pi^2 a_{1,1}$, 
$d_4\simeq -0.711003\equiv 8\pi^2 s_1^g$, 
and the truncated $f_2$ defined in Eq.(\ref{f2def}) reading explicitly
\be
f_2 = 1+\left(\frac{11}{2\pi} \alpha_s -\frac{19}{8\pi^2} \alpha_s^2\right)\ln \frac{m_E}{M_s}+ {\cal O}(\alpha_s^3) .
\ee
Re-expanding perturbatively Eq.(\ref{PNLLcompact}), one reproduces the NNLO pressure and $N^3LO$ ${\cal O}(\alpha_s^3)$ 
terms in Eqs.(\ref{Pas2}), (\ref{Psoftg3}), with modified coefficients in Eq.(\ref{p210mod}). 

\end{document}